\documentclass{article}
\newcommand{\bm}[1]{\mbox{\boldmath $#1$}}
\def\Nev{N_{\rm ev}}
\def\rmf{{f}}

\input epsf

\newcommand{\oeawpreprint}[2]
{
\noindent
\begin{minipage}[t]{\textwidth}
\begin{center}
\framebox[\textwidth]{$\rule[6mm]{0mm}{0mm}$ 
\raisebox{1.3mm}{Institut f\"ur Hochenergiephysik der \"Osterreichischen
Akademie der Wissenschaften}}

\vspace{2mm}    \rule{\textwidth}{0.2mm}\\
\vspace{-4mm}   \rule{\textwidth}{1pt}
\mbox{ }    #1    \hfill    #2   \mbox{ }\\
\vspace{-2mm}   \rule{\textwidth}{1pt}\\
\vspace{-4.2mm} \rule{\textwidth}{0.2mm}
\end{center}
\end{minipage}

}   

\begin{document}

\pagestyle{empty}

\oeawpreprint{February 1996}{HEPHY-PUB 639/96}

\vspace*{10mm}

\mbox{ }\hfill hep-ph/9604374

\vspace*{20mm}

\begin{center}
{\Large\bf Correlations, correlation integrals and 
application to Bose-Einstein interferometry\footnote{
Invited lecture given by PL at the Triangle Meeting: 
School and Workshop on Heavy Ion Collisons,
Bratislava, Slovakia,
4--9 September 1995.}
}\\

\mbox{ }\\
 P.\ Lipa, 
H.C.\ Eggers\footnote{
              Present address: Dept of Physics, University of
              Stellenbosch, 7600 Stellenbosch, South Africa.},
and B.\ Buschbeck \\
\mbox{ }\\
{\it 
     Institut f\"ur Hochenergiephysik der \"Osterreichischen
     Akademie der Wissenschaften,} \\
{\it 
     Nikolsdorfergasse 18, A--1050 Vienna, Austria}\\
\end{center}

\vspace*{0.9cm}

\begin{abstract}
We review the basic notion of correlations in point processes,
adapted to the language of high energy physicists. The measurement of
accessible information on correlations by means of correlation
integrals is summarized. Applications to the measurement of
Bose-Einstein interferometry as well as some pitfalls are discussed.
\end{abstract}

\vfill\noindent
\mbox{} \hfill (submitted to Acta.\ Phys.\ Slovakia)

\newpage
\pagestyle{plain}
\section{Introduction}

In moving to ever higher energies in particle physics, the
experimentalist faces a rapidily increasing complexity of the
observed events. Beyond posing questions of physics alone,
the task to compare experimental data with competing
theoretical and phenomenological models raises many questions 
purely {\em statistical\/} in nature. The
latter define a field of their own: {\em multiparticle statistics}.

The task of multiparticle statistics is to provide a framework 
for maximal utilization of the information provided by
experiments. This includes questions of measurement of single and
multiparticle spectra, correlations among two and more particles,
statistical errors due to finite event-sample size, limited detector
acceptance, misidentification of tracks and much more.

In these lectures, we summarize
basic concepts of point processes,  correlations  and their
measurement. To demonstrate the discriminative power of advanced
correlation measurements, we discuss a  recent test of
dynamical assumptions in modeling Bose-Einstein correlations.

\section{Cross sections, point processes and correlations}

Correlations (in the widest sense) among final state particles are
commonly expressed in terms of exclusive or
inclusive  differential cross sections denoted $\sigma_{\rm excl}$ and
$\sigma_{\rm incl}$ respectively.  To be specific, we recall some common
definitions \cite{Koba}:
\begin{eqnarray} \label{eq:aa1}
j_N(\bm{p}_1,\ldots,\bm{p}_N)
      & \equiv & 
		{1 \over \sigma_{\rm tot}}
  { d^{3N}\sigma_{\rm excl} \over 
  d^{3}\bm{p}_1\,d^{3}\bm{p}_2\cdots d^{3}\bm{p}_N} \;, 
\\  \label{eq:aa2}
\rho_q(\bm{p}_1,\ldots,\bm{p}_q)
      & \equiv & 
		{1 \over \sigma_{\rm tot}}
  { d^{3q}\sigma_{\rm incl} \over 
d^{3}\bm{p}_1\,d^{3}\bm{p}_2\cdots d^{3}\bm{p}_q} \;;
\end{eqnarray}
$\bm{p}_i$ is the three-momentum of the $i$-th particle,
and $\sigma_{\rm tot}=\sum_N \sigma_N$
is the total (inelastic) cross section, with
$\sigma_N$ the integrated cross section for events with 
$N$ final state particles (here to be taken as identical for
simplicity of notation). The ratio $P_N=\sigma_N/
\sigma_{\rm tot}$ gives the multiplicity distribution in full
phase-space.

The densities $j_N$ and $\rho_q$ are {\em very\/} different
objects: 
$j_N(\bm{p}_1,\ldots,\bm{p}_N) 
d^{3}\bm{p}_1\cdots d^{3}\bm{p}_N$
is proportional to the  probability that in an event with exactly
$N$ particles we find simultaneously the first particle in a box
of size  $d^{3}\bm{p}_1$ centered on 
$\bm{p}_1$, the second in a box at $\bm{p}_2$, \ldots, and 
the $N$-th at $\bm{p}_N$.
Thus,  $j_N$ characterizes the statistical
properties of samples built {\em exclusively\/} from events
with exactly $N$ particles (exclusive samples).
By contrast, $\rho_q(\bm{p}_1,\ldots,\bm{p}_q) 
d^{3}\bm{p}_1\cdots d^{3}\bm{p}_q$ is the {\em average number\/}
of unordered $q$-tuples of particles with momenta simultaneously
within infinitesimal boxes centered on $\bm{p}_1$,\ldots, $\bm{p}_q$
per event (no matter what $N$ is). Therefore  $\rho_q$ 
characterizes samples which include {\em all\/} events (inclusive
samples). While various other names are in use,
we follow Ref.\ \cite{DVJ88} in referring to $j_N$ as a {\em
Janossy density\/} and to $\rho_q$ as a {\em factorial moment
density\/}.

Both functions are symmetrized with respect to permutations of
their arguments. In fact, the labels ``first" particle, 
``second" particle,\ldots
are arbitrary and carry no physical information. Therefore
it is customary to count all $N!$ ($q!$) permutations of labels
as separate, independent events ($q$-tuples). This symmetrization
has nothing to do with quantum mechanical ``indistinguishability"
of particles, but reflects a mere convention designed to 
simplify the resulting formalism.

While 
$j_N/N!P_N$, normalized to unity, represents a conventional 
{\em joint probability density\/} of $N$ random variables $\bm{p}_i$,
it is a mistake to regard $\rho_q$ as such. It is constructed from
inclusive samples,  where the number of random variables ($N$) is a
again a random variable.

In fact, integration over  an arbitray region of phase-space
$\Omega$ gives the $q$-th factorial moment of the random
multiplicity $n$ in this domain:
\begin{equation} \label{eq:aat}
\xi_q(\Omega) = \int_\Omega \rho_q(\bm{p}_1,\ldots,\bm{p}_q)
\, d^3\bm{p}_1 \ldots d^3\bm{p}_q 
= \left\langle n^{[q]} \right\rangle_\Omega \;.
\end{equation} 
We use the common notation  
$n^{[q]} \equiv  n(n{-}1)\cdots(n{-}q{+}1)$
for factorials.

All the above definitions are given in terms of cross sections.
However, all {\em experimental\/} multiplicity and correlation
measurements can be viewed as particular {\em counting procedures\/}
of particles or particle-tuples in certain domains. Thus, we pose the
questions: How do we have to count in order to obtain a certain type
of information and, moreover, how do we count most efficiently?

The appropriate tool to tackle such questions is the 
theory of point
processes \cite{Srini74,DVJ88}. 
Consider a sample of $\Nev$ events, each labeled by an
index $a=1,\ldots,\Nev$. For greater generality, we subsequently
denote the ``positions" of the particles (=points) by 
$\bm{X}_i^a$, ($i=1,\ldots,N$), where $\bm{X}$ can refer to 
{\em any\/} set of coordinates. Frequently used examples are rapidity $y$, 
rapidity-azimuth $(y,\Phi)$ or even a combination of continuous and
discrete variables such as $(\bm{p},s)$, with $s$ labeling the
charge, spin or species of the particle\footnote{In this 
case all following integrations contain implicitly sums over $s$.}.

The density of such points at $\bm{x}$ in one particular event $a$
is most conveniently represented by the ``random Dirac comb" 
\begin{equation} \label{eq:aas} 
\hat\rho_1^a(\bm{x}) =
\sum_{i_1=1}^N \delta(\bm{x} - \bm{X}_{i_1}^a) \,, 
\end{equation}
which, when integrated over a certain domain $\Omega$, just gives the
number of particles $n$ in that domain. More generally, the
simultaneous behavior ( = correlation) of $q$ of these particles in
that event is represented by the restricted tensor product of Dirac
combs
\begin{equation} \label{eq:ciaa}
\hat\rho_q^a(\bm{x}_1,\ldots,\bm{x}_q) =  
\sum_{\scriptstyle i_1\neq
i_2\neq\ldots\neq i_q \atop \scriptstyle =1}^{N} \delta
(\bm{x}_1-\bm{X}^a_{i_1})\, \delta (\bm{x}_2-\bm{X}^a_{i_2})   
\,\cdots\, \delta (\bm{x}_q-\bm{X}^a_{i_q}) \,, 
\end{equation}
which just acts as a $q$-tuple counter in event $a$.
Note that it doesn't make sense to count the same particle more than
once (hence the restriction on the indices in the multiple sum). 

Meaningful results are extracted by averaging over the  inclusive
event sample, to yield the $q$-tuple density
\begin{equation} \label{eq:aau} 
\rho_q = \left\langle \hat\rho_q \right\rangle 
       = \Nev^{-1} \sum_{a=1}^{\Nev} \hat\rho_q^a \,,
\end{equation}
which is nothing but the counting prescription for
the factorial moment densities eq.~(\ref{eq:aa2}).

A point process is fully determined by the knowledge of
either all $\rho_q$, all $j_N$ or, most conveniently,
by its generating functional. The latter is  defined
by
\begin{eqnarray} \label{eq:gfa}
Z[\lambda(\bm{x})] & = &
\left\langle
\exp\left(
\int \hat\rho(\bm{x})\log\left[1+\lambda(\bm{x})\right]\,d\bm{x}
\right)
\right\rangle  
\\ \label{eq:gfb}
 & = &
\sum_{q\geq 0} \frac{1}{q!} \int 
\rho_q(\bm{x}_1,\ldots,\bm{x}_q)
\lambda(\bm{x}_1)\cdots\lambda(\bm{x}_q)\,
d\bm{x}_1\cdots d\bm{x}_q \;.
\end{eqnarray}
Once we know $Z[\lambda(\bm{x})]$,
the $\rho_q$ as well as the 
$j_N$ can be obtained via functional derivatives with respect to 
the test function $\lambda$:
\begin{eqnarray} \label{eq:gf1}
\rho_q(\bm{x}_1,\ldots,\bm{x}_q)
 & = &
 \left.
\frac{\delta^q Z[\lambda(\bm{x})]}
 {\delta\lambda(\bm{x}_1)\cdots\delta\lambda(\bm{x}_q)}
 \right|_{\lambda=0}
  \\ \label{eq:gf2}
j_N(\bm{x}_1,\ldots,\bm{x}_N)
 & = &
 \left.
\frac{\delta^q Z[\lambda(\bm{x})-1]}
 {\delta\lambda(\bm{x}_1)\cdots\delta\lambda(\bm{x}_N)}
 \right|_{\lambda=0} \;.
\end{eqnarray}

In order to understand what we mean by ``correlations"
 we first discuss the meaning of {\em
statistical independence\/} in point processes. The latter is
defined by full factorization  of the factorial moment 
densities $\rho_q(\bm{x}_1,\ldots,\bm{x}_q) = \rho_1(\bm{x}_1) 
\rho_1(\bm{x}_2) \cdots \rho_1(\bm{x}_q) $ for all $q$. In this case
we can readily sum up the series (\ref{eq:gfb}) to obtain
\begin{equation} \label{eq:gfp} 
Z[\lambda(\bm{x})]  = 
\exp\left(
\int\rho_1(\bm{x})\lambda(\bm{x})\,d\bm{x}
\right) \;,
\end{equation}
which is the generating functional of the so-called Poisson process.
In this process, the multiplicity of points in {\em any\/} arbitrary
domain $\Omega$ follows a Poisson distribution.
This process plays a similar central role for point processes 
as the Gaussian does for the statistics of continuous random variables.

Genuine correlations among points are then quantifiable by
{\em deviations\/} from the Poisson process. Consider
the family of functions, called {\em cumulant densities\/},
obtained by functional derivatives of $\log Z[\lambda]\,$:
\begin{equation} \label{eq:gf3} 
C_q(\bm{x}_1,\ldots,\bm{x}_q)
  = 
 \left.
\frac{\delta^q\, \log Z[\lambda(\bm{x})]}
 {\delta\lambda(\bm{x}_1)\cdots\delta\lambda(\bm{x}_q)}
 \right|_{\lambda=0} \;.
\end{equation}
For the Poisson process (\ref{eq:gfp}) we see that $C_1=\rho_1$ and 
all higher cumulant densities vanish. 
In this sense, nonvanishing $C_q$ with $(q\geq 2)$ quantify
genuine correlations, i.e. deviations from the Poisson process.

It is quite possible that only a few orders of cumulants
are nonzero. For example in heavy ion reactions cumulants of third
and higher orders are highly suppressed. 
Such processes can be modeled as 
poisson processes of ``clusters", each cluster decaying into one
or 2 daughter particles.

The generating functional $Z[\lambda]$ is a convenient bookkeeping
device for establishing all kinds of relations among $j_N$, $\rho_q$
and $C_q$, for example
\begin{eqnarray} \label{eq:cum2} 
C_2(\bm{x}_1,\bm{x}_2) &=& \rho_2(\bm{x}_1,\bm{x}_2) - 
\rho_1(\bm{x}_1)\rho_1(\bm{x}_2) \,, 
\\ \label{eq:cum3}
C_3(\bm{x}_1,\bm{x}_2,\bm{x}_3) &=& \rho_3(\bm{x}_1,\bm{x}_2,\bm{x}_3) 
               -\ \rho_1(\bm{x}_1)\rho_2(\bm{x}_2,\bm{x}_3)
               -\ \rho_1(\bm{x}_2)\rho_2(\bm{x}_3,\bm{x}_1)\nonumber\\
    & &\mbox{} -\ \rho_1(\bm{x}_3)\rho_2(\bm{x}_1,\bm{x}_2)
            +\ 2\,\rho_1(\bm{x}_1)\rho_1(\bm{x}_2)\rho_1(\bm{x}_3)
                 \quad \mbox{etc.}
\end{eqnarray}
Notice the subtraction of various products of lower order 
factorial moment densities in the construction of cumulants,
 sometimes called ``removal of combinatorial background". It is this 
 property that makes cumulants a favourable tool
 for discriminative experimental anlysis, but at the same time
 more difficult to measure.

\section{Correlation integrals}

If we had experimental knowledge of correlations in terms of 
either $j_N$,
$\rho_q$ or $C_q$ to {\em all\/} orders, the multiparticle 
production process would be {\em completely\/} determined. 
This can never, of course, be achieved in practice. However,
while higher-order
correlation functions can never be sampled fully differentially, 
one can still try to  extract as much information as possible from
a give data sample by measuring
{\em integrals} over various domains $\Omega$ as in eq.\
(\ref{eq:aat}).

Conventional measurements of correlations proceed first to discretize
the continuous variable $\bm{X}$ (``binning the data``) and then
to find $\xi_q$ by averaging over all events the counts 
$n_m^{[q]}$ in every bin \cite{Bia86a},
\begin{equation} \label{eq:} 
\xi_q^{\rm conv} = \left\langle \sum_{{\rm bins}\ m} n_m^{[q]} \right\rangle .
\end{equation}
By contrast, so-called {\it correlation integrals}
rely on distances between pairs of points  $X_{i_1 i_2} \equiv |
\bm{X}_{i_1} - \bm{X}_{i_2} |$  rather than counting particles in
predefined bins \cite{Bia86a,Lip92a}. More generally, every
correlation integral of order $q$ assigns a ``size" 
to every possible
$q$-tuple of particles. The way this assignment is done distinguishes
the different versions of correlation integrals. Finally, they count
the number of $q$-tuples with a given size $\epsilon$ 
(``differential forms")
or the ones smaller than a given size (``integral forms"). For
a large data sample, this corresponds to an integration over $\rho_q$
in specific domains $\Omega(\epsilon)$.

For the Star integral \cite{Egg93a}, the domain
$\Omega$ is given by the collection of $N$ spheres of radius $\epsilon$,
each  centered at one of the  $N$ particles in the event. For a given
event $a$ the number of particles (``sphere count'') within one of these
spheres is, not counting the particle at the center $\bm{X}_{i_1}$,
\begin{equation} \label{eq:ciad}  
 a \equiv 
\hat n(\bm{X}_{i_1},\epsilon) 
\equiv 
\sum_{i_2=1}^N \Theta(\epsilon - X_{i_1 i_2})
\,,\ \ \ \ \ \ i_2 \neq i_1 \,,
\end{equation}
where $a$ is an ``ultra short'' notation needed for some lengthy formulae
below. With this elementary counter the factorial moment of order $q$ is
simply obtained by
\begin{equation} \label{eq:ciac}
\xi_q^{\rm Star}(\epsilon) = \left\langle
\sum_{i_1} 
\hat n(\bm{X}_{i_1},\epsilon)^{[q-1]} \right\rangle
= \left\langle
\sum_{i_1} 
a^{[q-1]} \right\rangle.
\end{equation}
One can show \cite{Egg93a} that
the above counting prescription corresponds to integrating 
eq.~(\ref{eq:aat})
using for $\Omega$ a particular ``Star'' domain implemented via theta
functions $\Theta_{1j} \equiv \Theta(\epsilon - |\bm{x}_1 - \bm{x}_j|)$,
restricting all $q{-}1$ coordinates $\bm{x}_j$ to within a distance
$\epsilon$ of $\bm{x}_1$: 
\begin{equation} \label{eq:abba} 
\xi_q^{\rm Star}(\epsilon) =
\int \rho_q(\bm{x}_1,\ldots,\bm{x}_q) \,
\Theta_{12}\Theta_{13}\ldots\Theta_{1q} \,
d\bm{x}_1 \ldots d\bm{x}_q \, .
\end{equation}

The superiority of  correlation integrals in general over the
conventional  Bia\l as-Peschanski factorial moments \cite{Bia86a} is
discussed and demonstrated in \cite{Lip92a}. The further advantage of the
particular ``Star" domain  lies in the fact that eq.~(\ref{eq:ciac}) requires
typically $\Nev*N^2$ computation steps for {\em any\/} order $q$, whereas
other types of correlation integral have order $\Nev*N^q$ complexity,
which quickly becomes unmanageable for higher orders. 

In order to eliminate, among other things, the overall total cross
section, it has become customary in high energy physics to measure 
{\it normalized factorial moments} \cite{Bia86a}. The denominator 
used for such normalization should be made up of the 
{\em uncorrelated\/}
background, $\rho_1^q$. While it can be implemented in a number of 
ways, we prefer the ``vertical'' normalization, in which $\rho_1^q$ 
is integrated over exactly the same domain $\Omega$ as the inclusive 
density $\rho_q$ in the numerator. Thus for the Star integral, the 
normalized moment is 
\begin{equation} \label{eq:cigc}
F_q^{\rm Star}(\epsilon) \equiv
{\xi_q^{\rm Star} \over \xi_q^{\rm norm}  }  = 
{
\int \rho_q(\bm{x}_1,\ldots,\bm{x}_q) \,
\Theta_{12}\Theta_{13}\ldots\Theta_{1q} \,
d\bm{x}_1 \ldots d\bm{x}_q
\over
\int \rho_1(\bm{x}_1)\ldots\rho_1(\bm{x}_q) \,
\Theta_{12}\Theta_{13}\ldots\Theta_{1q} \,
d\bm{x}_1 \ldots d\bm{x}_q
} \,.
\end{equation}
We have shown \cite{Egg93a} that 
the denominator $\xi_q^{\rm norm}$ is given by the following double
event average: with  $X_{i_1 i_2}^{ab} \equiv
| \bm{X}_{i_1}^{a} - \bm{X}_{i_2}^{b} |$ measuring the distance 
between two particles {\it taken from different events}
$a$ and $b$, and the ``ultra short" notation
\begin{equation} \label{eq:evda}
b \equiv 
\hat n_b(\bm{X}_{i_1}^a,\epsilon) = 
\sum_{i_2} \Theta(\epsilon - X_{i_1 i_2}^{ab})
\end{equation}
we get
\begin{equation} \label{eq:evd}
\xi_q^{\rm norm}(\epsilon) \equiv 
\left\langle \sum_{i_1} 
\hat\xi_q^{\rm norm}(i) \right\rangle =
\left\langle \sum_{i_1} 
\left\langle b  \right\rangle^{q-1} \right\rangle \;.
\end{equation}
Note that the outer event average and sum over $i_1$ are taken over
the center  particle taken from event $a$, each of which is used as
the center of sphere counts $\hat n_b(\bm{X}_{i_1}^a,\epsilon)$
taken over other  events $b\neq a$ in the inner event average. We
thus see the natural  emergence of the heuristic procedure of
normalization known as ``event mixing'' \cite{Egg93a,Lip92a}. This
counting prescription has complexity $\Nev^2*N^2$ --- unmangeable
for large event samples. However, in practical applications it is
sufficient to reduce the inner average over $b$-events to a small
subsample
$b=a-1,a-2,\ldots,a-A$, consisting only of $A-1$ events and requiring
only $A*\Nev*N^2$ computation steps. This procedure we call
``reduced" event mixing.

Another essential advantage of the Star domain is the fact that 
on top of the $F_q^{\rm Star}$  we get the integrals of cumulants
(almost) for free, once the counts $a$ and $b$ per particle $i$ are
perfomed. Integrating the $C_q$ over the Star integral domain, 
we define normalized (factorial) cumulants 
$
K_q^{\rm Star}(\epsilon) \equiv 
{ \rmf_q(\epsilon) / \xi_q^{\rm norm}(\epsilon) } 
$,
with
\begin{equation} \label{eq:cui}
\rmf_q(\epsilon) \equiv 
\int C_q(\bm{x}_1,\ldots,\bm{x}_q) \,
\Theta_{12}\Theta_{13}\ldots \Theta_{1q} \,
d\bm{x}_1\ldots d\bm{x}_q \ .
\end{equation}
The latter can be written
entirely in terms of the sphere counts $a$ and $b$ introduced previously!
Defining for convenience the ``$i$-particle cumulant''
$\hat\rmf_q(i)$ so that
$
\left\langle \sum_i \hat\rmf_q(i) \right\rangle = \rmf_q \,,
$
we find \cite{Egg93a}
\begin{eqnarray}
\label{eq:cupa}
\hat\rmf_2(i) &=& a - \langle b \rangle \,, \\
\label{eq:cupb}
\hat\rmf_3(i) 
&=& a^{[2]} - \langle b^{[2]} \rangle - 2 a \langle b \rangle
            + 2 \langle b \rangle^2 \ , \quad \mbox{etc.}  
\end{eqnarray}

While computationally more expensive than the Star integral by 
orders of magnitude, 
other types of integration domains $\Omega$ are also useful, 
in particular when 
a comparison of theoretical models with data
dictate a specific choice of  variables and integration domain.
This is typically the case in
measurements of Bose-Einstein correlations, where a preferred
variable is the 4-momentum difference 
$q_{ij} = [(\bm{p}_i - \bm{p}_j)^2 - (E_i - E_j)^2]^{1/2}$.

We then may study differential forms of integrations over
$\rho_q$ or $C_q$ in a fully symmetrical manner (called GHP 
topology) \cite{Egg93d} such as
\begin{equation} \label{eq:abb} 
C_q(\epsilon) =
\int C_q(\bm{p}_1,\ldots,\bm{p}_q) \,
 \delta(\epsilon - \sum_{i<j=1}^q q_{ij}^2)\,
d^3\bm{p}_1 \cdots d^3\bm{p}_q \, .
\end{equation}
Other $q$-tuple size prescriptions (used below) are obtained
by changing the argument of the $\delta$-function appropriately,
e.g.\ $\delta(\epsilon - \sum_{i<j=1}^q q_{ij})$ or
$\delta(\epsilon - \max(q_{12},\ldots,q_{q-1,q}))$.

The corresponding counting prescriptions are conveniently
written in terms of the generic $q$-tuple counter
\begin{equation} \label{eq:qcntr} 
I^{e_1e_2\cdots e_q}_{i_1i_2\cdots i_q}(\epsilon) =
\left\{ \begin{array}{ll}
         1 & \mbox{if the ``size" of the $q$-tuple is within 
         $\epsilon+\delta\epsilon$} \\
         0 & \mbox{otherwise,}
         \end{array}
\right.
\end{equation}
where the $q$-tuple is composed of particle $i_1$ taken from
event $e_1$, particle $i_2$ taken from a different event $e_2$, 
etc., and the ``size" is evaluated, e.g., according to one of above
quoted prescriptions.

The counting algorithms for $\rho_q(\epsilon)$ and the first 
$C_q(\epsilon)$ 
are
\begin{eqnarray} \label{eq:cicnt1} 
\rho_q(\epsilon)\delta\epsilon & = &
\left\langle
\sum_{i_1\neq\cdots\neq i_q} I^{a a\cdots a}_{i_1 i_2\cdots i_q}
(\epsilon)
\right\rangle
\\ \label{eq:cicnt2} 
C_2(\epsilon)\delta\epsilon & = &
\left\langle
\sum_{i\neq j} I^{a a}_{i j}(\epsilon)
\right\rangle_{\!\!\! a}
-
\left\langle\left\langle
\sum_{i,j} I^{a b}_{ij}(\epsilon)
\right\rangle_{\!\!\! b}\right\rangle_{\!\!\! a}
\\ \label{eq:cicnt3} 
C_3(\epsilon)\delta\epsilon & = &
\left\langle
\sum_{i\neq j\neq k} I^{a a a}_{i j k}(\epsilon)
\right\rangle_{\!\!\! a}
-
3\left\langle\left\langle
\sum_{i\neq j,k} I^{a a b}_{i j k}(\epsilon)
\right\rangle_{\!\!\! b}\right\rangle_{\!\!\! a}
+
2\left\langle\left\langle\left\langle
\sum_{i,j,k} I^{a b c}_{i j k}(\epsilon)
\right\rangle_{\!\!\! c}\right\rangle_{\!\!\! b}\right\rangle_{\!\!\! a} \;.
\end{eqnarray}

Integrals (\ref{eq:abb}) over uncorrelated tensor products of 
$\rho_1$, needed for normalization,
are sampled in similar ways \cite{Egg93d}:
\begin{equation} \label{eq:nrmcnt} 
\rho_1{\otimes}\rho_1{\otimes}\cdots{\otimes}\rho_1(\epsilon)
\delta\epsilon  = 
\left\langle\left\langle\cdots\left\langle
\sum_{i_1,i_2,\cdots,i_q} I^{e_1 e_2\cdots e_q}_{i_1 i_2
\cdots i_q}(\epsilon)
\right\rangle_{\!\!\! e_1}\right\rangle_{\!\!\! e_2}
                    \cdots\right\rangle_{\!\!\! e_q}
\;.
\end{equation}
Note that multiple event averages, e.g.\   
$\langle\langle \ldots \rangle_a\rangle_b= 
\sum_{a\neq b}/\Nev(\Nev-1)$, always run over unequal events
to avoid noticeable sampling biases.
Again, inner event averages can  run over a small
fraction of the full sample to keep computing 
times in manageable ranges \cite{Egg93a}.

\section{Applications to Bose-Einstein correlation measurements}

As an application, we discuss a recent test \cite{Egg95a}
of quantum statistical (QS) models 
 for Bose-Einstein correlations among
 identical pions \cite{Biy90a,Plu92a,And93a}. 
 The latter postulate a specific form
 of the generating functional (\ref{eq:gfa}) with $C_2$
 as freely parametrizable function. Once $C_2$ is
 determined from experiment, all higher cumulant densities
 are fully specified with no further adjustable parameter.
 One therefore can test experimentally the validity of
 the postulated generating functional by comparing higher
 order cumulants from experiment with the model predictions,
 without having to rely on a particular parametrization of $C_2$.

 When relative
phases are neglected, the second and third (reduced) ``QS cumulants'' 
of interest are \cite{And93a}
\begin{eqnarray}
\label{eq:cue}
k_2 \equiv {C_2 \over \rho_1{\otimes}\rho_1}
   &=& 2p(1-p)d_{12} + p^2 d_{12}^2 \,, \\
\label{eq:cuf}
k_3 \equiv {C_3 \over \rho_1{\otimes}\rho_1{\otimes}\rho_1}
   &=& 2p^2(1-p)[ d_{12}d_{23} + d_{23}d_{31} + d_{31}d_{12} ]
      + 2p^3 d_{12} d_{23} d_{31} \,,
\end{eqnarray}
where $\rho_1{\otimes}\rho_1 = \rho_1(\bm{p}_1)\rho_1(\bm{p}_2)$
and the $d_{ij} {=} d(q_{ij})$ are functions of the 4-momentum
differences 
$q_{ij}$.

While in principle calculable from a given density matrix, the
functions $d_{ij}$ are usually parametrized in a plausible and/or
convenient way, such as Gaussian ($d_{ij} = \exp(-r^2 q_{ij}^2) $), 
exponential ($d_{ij} = \exp(-r q_{ij}) $) and power-law
parametrizations ($d_{ij} = q_{ij}^{-\alpha}$). Since the Gaussian
does not give viable fits, only the latter two parametrizations 
are used below.

\begin{figure}[th]
\centerline{
\epsfxsize=135mm
\epsfbox{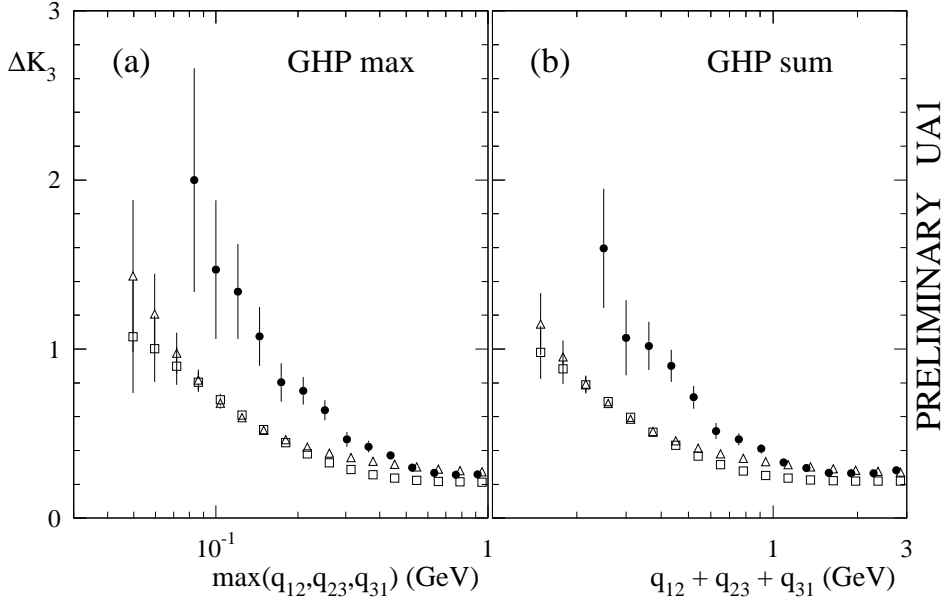}
}
\caption{
Third-order differential cumulants $\Delta K_3(\epsilon)$ integrated
with  GHP max (a) and GHP sum (b) topologies \protect\cite{Egg95a}. 
Filled circles
represent  UA1 minimum-bias data, while open symbols show
predictions for $\Delta K_3$ from QS theory and parameter values
obtained from fits to $\Delta K_2$.  Open triangles are predictions
based on the QS power-law parametrization; open squares are QS
exponential predictions. }
\label{fig1}
\end{figure}

In Figure 1, we show the results of a preliminary\footnote{ This
preliminary analysis is based on like-sign particles in a 
restricted  phase space region with good detector acceptance. The
analysis is currently being extended to an enlarged phase space region. }
comparison of the  normalized third order cumulant $\Delta K_3$
(\ref{eq:cicnt3}) obtained from 
  UA1 minimum-bias data ($\bar p p$-reactions at $\sqrt{s}=
630\,$GeV) with QS-predictions (\ref{eq:cuf}). Besides the
GHP-sum topology used in (b), we show in (a) a separate analysis
using the ``GHP max'' topology \cite{Egg93a}, which bins triplets
according to the largest of the three momentum differences,
max$(q_{12},q_{23},q_{31})$. Fit parameter values used for the
respective power law and exponential parametrizations were taken
from the  QS fit (\ref{eq:cue}) to $\Delta K_2$ obtained from the same
data sample.  All theoretical points shown are determined only up to
an additive constant, so that the curves may be shifted up and down.
It is clear, though, that the shape of third-order cumulant
data measured differs appreciably from that predicted by the QS
formulae and parameter values from $\Delta K_2$. This conclusion
holds independently of the topology used and of the functional form
taken for $d$.

The results of this analysis may appear, at first sight, to
contradict the conclusion \cite{Plu92a}, based on an earlier UA1
paper \cite{UA1-92a},  that QS theory was compatible with
higher-order {\em factorial moments}. The apparent discrepancy is 
explained by pointing out that the recent improvement of
measurement techniques have permitted
the present direct measurements of {\em cumulants}, which are
considerably more  sensitive than moments. The latter are
dominated numerically by the combinatorial background of lower order 
correlations and thus contain mostly redundant information.

\section{Final remarks}
 
The use of correlation integrals permits much more
accurate measurements and hence will likely reveal more detailed
structure of the underlying dynamics. 
In particular, cumulants are promising and sensitive tools to
obtain refined insight into various production mechanisms. 

Greater accuracy requires,
however, that possible biases be understood on a deeper level than
before. One such bias arising generally in the measurement of
correlations is due to the {\em finite\/} size of  event samples and
can have quite noticeable size. In practice, one has to use {\em
unbiased\/} estimators, which correct for this effect (cf.\
\cite{Egg93a}). 

Another caveat is the comparison of  experimental
correlation data with theoretical predictions, when the latter
are given in terms of differentially normalized (reduced)
correlation functions such as (\ref{eq:cue}) and (\ref{eq:cuf}). 
{\it Experimentally\/}, one can never measure fully
differential ratios; rather, the numerator and denominator
are averaged over some bin of finite size $\Omega$ (however small)
before the ratio is taken. The discrepancy to
fully differential ratios can be quite substantial.
A procedure to overcome this problem amounts to a
Monte Carlo integration of a {\it theoretical correlation function\/}
sampled according to the {\it experimental uncorrelated one-particle
distribution\/}; this and other details are
explained in \cite{Egg95a}.
\\

{\bf Acknowledgements}: This work was supported by the Austrian
Academy of Science by means of an APART (Austrian Programme for
Advanced Research and Technology) fellowship (PL) and by the
Austrian Fonds zur F\"orderung der Wissenschaften (FWF) by means of
a Lise-Meitner fellowship (HCE).



\begin{thebibliography}{99}
\parskip=0cm
\itemsep=0cm

\bibitem{Koba}
Z.\ Koba, Acta Phys.\ Pol.\ {\bf B4}, 95 (1973).

\bibitem{DVJ88}
D.J.\ Daley and D.\ Vere-Jones, {\it An Introduction to the Theory
of Point Processes}, Springer, New York, 1988.

\bibitem{Srini74}
S.K.\ Srinivasan, {\it Stochastic Point Processes and their Applications},
Griffin's statistical monographs and courses, no. 34, Hafner Press,
 New York, 1974.
 
\bibitem{Bia86a}A.\ Bia\l as and R.\ Peschanski, Nucl.\ Phys.\ 
{\bf B273}, 703 (1986); {\bf B308}, 857 (1988).


\bibitem{Lip92a}P.\ Lipa et al.,
           Phys.\ Lett.\ {\bf B285}, 300 (1992).

\bibitem{Egg93a}H.C.\ Eggers et al., 
           Phys.\ Rev.\ {\bf D48}, 2040 (1993);
           Phys.\ Rev.\ {\bf D51}, 2138 (1995).

\bibitem{Egg93d}H.C.\ Eggers et al.,
           Phys.\ Lett.\ {\bf B301}, 298 (1993).

\bibitem{Egg95a}
         H.C.\ Eggers, B.\ Buschbeck and P.\ Lipa, HEPHY-PUB 634/95,
         to be published in the Proceedings of the XXV-th 
         International Symposium on Multiparticle Dynamics, Star\'a
         Lesn\'a, Slovakia, 12-16 Sept.\ 1995.

\bibitem{Biy90a}M.\ Biyajima, A.\ Bartl, T.\ Mizoguchi, N.\ Suzuki
           and O.\ Terazawa, Progr.\ Theor.\ Phys.\ 
           {\bf 84}, 931 (1990); {\bf 88}, 157A (1992).

\bibitem{Plu92a}M.\ Pl\"umer, L.V.\ Razumov and
           R.M.\ Weiner, Phys.\ Lett.\ {\bf B286}, 335 (1992).

\bibitem{And93a}I.V.\ Andreev, M.\ Pl\"umer, and R.M.\ Weiner, 
           Int.\ J.\ Mod.\ Phys.\ {\bf A8}, 4577 (1993).

\bibitem{UA1-92a}UA1 Collaboration, N.\ Neumeister {\it et al.}, 
            Phys.\ Lett.\ {\bf B275}, 186 (1992).
\end{thebibliography}
\end{document}